\documentclass[
floatfix,aps,prd,amsmath,amssymb,nofootinbib,preprintnumbers,superscriptaddress]{revtex4}

\usepackage{graphicx}
\usepackage{amsmath, amssymb, color}
\usepackage{mathrsfs}

\begin{document}

\preprint{ APCTP Pre2015-007, YITP-15-61, RESCEU-22/15}

\title{
Reheating signature in the gravitational wave spectrum from
self-ordering scalar fields
}
\author{Sachiko Kuroyanagi}
\affiliation{Asia Pacific Center for Theoretical Physics,
Pohang, Gyeongbuk 790-784, Korea}
\affiliation{Department of Physics, Nagoya University, Chikusa, Nagoya 464-8602, Japan}
\author{Takashi Hiramatsu}
\affiliation{Yukawa Institute for Theoretical Physics, Kyoto
University, Kyoto 606-8502, Japan}
\author{Jun'ichi Yokoyama}
\affiliation{Research Center for the Early Universe (RESCEU), 
School of Science, The University of Tokyo,
     Tokyo 113-0033, Japan }
\affiliation{Department of Physics, School of Science, The University of Tokyo,
Tokyo 113-0033, Japan  }
\affiliation{Kavli Institute for the Physics and Mathematics of the Universe (Kavli
IPMU), The University of Tokyo, Chiba, 277-8568, Japan}

\newcommand{\pd}[2]{\frac{\partial #1}{\partial #2}}
\newcommand{\pdd}[2]{\frac{\partial^2 #1}{\partial #2^2}}
\newcommand{\hh}{\chi_{ij}}
\newcommand{\tB}{\widetilde{B}_{ij}}
\newcommand{\tS}{\widetilde{S}_{ij}}
\newcommand{\kk}{\mbox{\boldmath $k$}}
\newcommand{\qq}{\mathbf{q}}
\newcommand{\xx}{\mbox{\boldmath $x$}}
\newcommand{\yy}{\mbox{\boldmath $y$}}
\newcommand{\fhh}{\widetilde{\chi}_{ij}}
\newcommand{\dhh}{\dot{\chi}_{ij}}
\newcommand{\dfhh}{\dot{\widetilde{\chi}}_{ij}}
\newcommand{\Mpl}{M_{\rm pl}}
\newcommand{\mpl}{m_{\rm pl}}
\newcommand{\HH}{\mathcal{H}}
\newcommand{\Lag}{\mathscr{L}}
\newcommand{\Ham}{\mathscr{H}}
\newcommand{\vol}[1]{\langle#1\rangle}

\newcommand{\red}[1]{{\color{red}#1}}

\begin{abstract}
  We investigate the imprint of reheating on the gravitational wave
  spectrum produced by self-ordering of multi-component scalar fields
  after a global phase transition.  The equation of state of the
  Universe during reheating, which usually has different behaviour
  from that of a radiation-dominated Universe, affects the evolution
  of gravitational waves through the Hubble expansion term in the
  equations of motion.  This gives rise to a different power-law
  behavior of frequency in the gravitational wave spectrum.  The
  reheating history is therefore imprinted in the shape of the
  spectrum.  We perform $512^3$ lattice simulations to investigate how
  the ordering scalar field reacts to the change of the Hubble
  expansion and how the reheating effect arises in the spectrum.  We
  also compare the result with inflation-produced gravitational waves,
  which has a similar spectral shape, and discuss whether it is
  possible to distinguish the origin between inflation and global
  phase transition by detecting the shape with future direct detection
  gravitational wave experiments such as DECIGO.
\end{abstract}

\maketitle

\section{Introduction}

Detection of gravitational waves is an exciting new observational
frontier in astrophysics and cosmology.  Ground-based laser
interferometric detectors of new generation, such as Advanced-LIGO
\cite{Harry:2010zz}, Advanced-VIRGO \cite{Accadia:2011zzc} and KAGRA
\cite{Somiya:2011np}, are currently under construction.  They promise
to yield new insights on many types of astrophysical events.  In
future, satellite experiments such as eLISA
\cite{AmaroSeoane:2012km,AmaroSeoane:2012je}, DECIGO
\cite{Seto:2001qf,Kawamura:2011zz} and BBO \cite{bbo} would enable us
to explore the Universe with an unprecedented sensitivity at lower
frequencies and provide a wealth of important information not only for
astronomy, but also for cosmology.

Thanks to the weak interactions with matter, gravitational waves offer
a unique opportunity to directly observe the earliest epochs of the
Universe beyond the last scattering surface of photons.  One
interesting source of gravitational waves in the early Universe is a
scalar field whose non-vanishing expectation value breaks a global
$O(N)$ symmetry \cite{Krauss:1991qu}.  After the phase
transition, the self-ordering of the Goldstone modes continuously
sources gravitational waves at the horizon scale and produce a
scale-invariant spectrum.

Such scale-invariant gravitational waves can be tested by various
types of experiments at different frequency bands.  While ground-based
direct detection experiments have sensitivity at $\sim 100$Hz, space
missions explore gravitational waves at lower frequencies; eLISA will
probe $\sim 10^{-3}$Hz; and DECIGO/BBO is designed to measure those at $\sim
0.1-1$Hz.  There are also indirect means of observations such as
B-mode polarization in the Cosmic Microwave Background (CMB)
\cite{Kamionkowski:1996zd,Zaldarriaga:1996xe,Planck:2006aa,Hanson:2013hsb,Ade:2014afa}
and pulsar timing observations
\cite{Hobbs:2009yy,vanHaasteren:2011ni,Demorest:2012bv,Manchester:2012za},
which can probe gravitational waves at $\sim 10^{-18}$Hz and $\sim
10^{-8}$Hz, respectively.

Several analytical and numerical studies have been conducted to
estimate the amplitude of the gravitational wave background from a
global phase transition
\cite{JonesSmith:2007ne,Fenu:2009qf,Giblin:2011yh,Figueroa:2012kw} and
their results are mutually consistent in the large $N$ limit.  The
effects on CMB have also been studied in the literature, in terms of
temperature and polarization anisotropies
\cite{Durrer:1998rw,GarciaBellido:2010if,Fenu:2013tea},
non-Gaussianity \cite{Adshead:2009bz,Figueroa:2010zx} and spectral
distortions \cite{Amin:2014ada}.  The difference from the inflationary
gravitational wave background, which also has a nearly scale-invariant
spectrum, has been studied in
\cite{Baumann:2009mq,Adshead:2009bz,Krauss:2010df} focusing on the CMB
scale.  Recently, the Planck satellite has placed an upper limit on
the effective defect energy scale of global textures, $G\mu<1.1\times
10^{-6}$ \cite{Ade:2013xla}, where $G$ is Newton's constant.  The
energy scale $\mu$ is related with the vacuum expectation value of the
$O(4)$ scalar fields, $v$, as $\mu=\pi v^2$. \footnote{ Note that the
  relation between $\mu$ and $v$ is different from the Planck paper by
  a factor of $2$, because we adopt a Lagrangian of real scalar fields
  in this paper, while the Planck paper uses the Lagrangian of complex
  scalar fields.  For details see the appendix in
  Ref. \cite{Urrestilla:2007sf}.}  In terms of $v$, the constraint is
rewritten as $v/m_{\rm pl}<6\times 10^{-4}$ with $m_{\rm
  pl}=G^{-1/2}=1.2\times 10^{19}$ GeV being the Planck scale.  The
contribution on the B-mode polarization signal has also been studied
motivated by the BICEP2 experiment \cite{Dent:2014rga,Durrer:2014raa}
and a slightly better constraint is obtained, $G\mu<7.3\times 10^{-7}$
\cite{Lizarraga:2014xza}, which corresponds to $v/m_{\rm pl}<5\times
10^{-4}$.  Since the peak of the signal in the B-mode spectrum arises
at different scale, the constraint on gravitational waves from a
global $O(N)$ phase transition is weaker compared to that of
inflation.

The purpose of this paper is to investigate the effect of reheating on
the gravitational wave spectrum of a global $O(N)$ phase transition,
and make a prediction for future direct detection experiments.  It is
known that the Hubble expansion rate of the Universe makes difference
in the power of generated gravitational waves
\cite{JonesSmith:2007ne,Krauss:2010df}.  It also affects the decline
rate of the amplitude due to redshift after generation.  After the
phase transition, gravitational waves are continuously generated as
the random initial configuration of the scalar field is homogenized up
to the Hubble horizon scale at each epoch, and information of
expansion rate is contained in the spectral amplitude of the mode.  In
the ordinary scenario of reheating, the Universe is dominated by the
coherently oscillating inflaton field with a quadratic inflaton
potential, which results in the same expansion rate as that of the
matter-dominated Universe.  When reheating is completed, the Universe
becomes radiation-dominated.  Therefore, information on the transition
of the expansion rate at the end of reheating may be imprinted in the
gravitational waves generated around the completion of reheating.  The
corresponding frequency of the gravitational wave spectrum is related
to the energy scale of reheating.

A similar effect arises in the spectrum of the inflationary
gravitational wave background.  Interestingly, it has been shown that
the effect arises in the sensitivity range of DECIGO and BBO, if the
energy scale of reheating, or similar mechanism to yield a
matter-dominated early Universe such as a late-time entropy
production, is around $10^7$GeV
\cite{Seto:2003kc,Boyle:2005se,Nakayama:2008ip,Nakayama:2008wy,Nakayama:2009ce,Kuroyanagi:2010mm,Kuroyanagi:2011fy,Kuroyanagi:2013ns}.
The same would be expected in the case of a global $O(N)$ phase
transition, and could be explored by those experiments.

In this paper, we perform lattice simulations to explicitly evaluate
the effect of reheating on the gravitational wave spectrum.  While
inflationary gravitational waves are generated when each mode crosses
outside the Hubble horizon during inflation, in the case of the $O(N)$
phase transition, gravitational waves are sourced by the anisotropic
stress of the scalar field when the mode enters the horizon after the
phase transition.  Because of the difference in the generation
process, the effect of reheating arises in a different way.  To obtain
the gravitational wave spectrum, we follow the evolution of both the
scalar field and gravitational wave in a 3-dimensional lattice with
changing the background equation of state.  Lattice simulations
provide an accurate estimate including non-linear dynamics of the
scalar field.

The paper is organized as follows.  In the next section, we describe
background equations during reheating, and the evolution equations of
the scalar field and gravitational waves, used in the lattice
simulations.  In Sec. \ref{sec:result}, we show the spectra obtained
from our simulations.  Then we discuss the differences between the
spectrum from the $O(N)$ phase transition and that from inflation.
Furthermore, we perform a Fisher matrix analysis and investigate
whether we can determine the reheating temperature by observing the
reheating signature in the spectrum with future experiments such as
DECIGO and Ultimate DECIGO.  We also discuss if we can determine the
origin of the observed gravitational wave background by measuring the
small differences between inflation and $O(N)$ phase transition
origin.  Section \ref{sec:conclusion} is devoted to conclusion.

\section{Model}

In this section, we describe setup for our simulation.  For the
background, we assume the conventional reheating scenario in which the
inflaton energy is transferred to the radiation energy while the
inflaton oscillates around the bottom of its quadratic potential.  The
inflaton energy is the dominant component during the reheating phase
and the expansion rate is the same as the matter-dominated Universe.
As the energy-transfer process proceeds, the Universe becomes
radiation-dominated.

Under this background, we consider a phase transition, where the
global $O(N)$ symmetry of a scalar field is broken to $O(N-1)$.  The
field rolls down to the true vacuum when the symmetry is broken, and
each causally disconnected region of the Universe gets arbitrarily
different directions of the field, which yields the spatial gradient
of the field on superhorizon scales.  As the comoving Hubble horizon
grows, previously causally disconnected regions come into contact and
the field moves to match the orientation, which is called
self-ordering.  The field releases gradient energy as it relaxes and
generates anisotropic stresses that source gravitational waves.

We consider the case where the symmetry breaking occurs well before
the universe becomes radiation dominated, so that the information on
the transition from matter- to radiation-dominated phase is imprinted
on the gravitational wave spectrum.

\subsection{Background equations for the reheating process}
We work in a spatially flat Friedmann-Lema{\^\i}tre-Robertson-Walker
background with the metric
%
\begin{equation}
 ds^2 = a^2(\tau)(-d\tau^2 + \delta_{ij}dx^idx^j),
\end{equation}
%
where $\tau$ denotes conformal time and $a(\tau)$ is the scale factor.
For the reheating process, we assume a perturbative decay of the
inflaton field $\varphi$ into radiation.  Assuming that the inflaton
potential $U$ is approximated as $U=m^2\varphi^2/2$ during the
oscillating phase, the equation for the energy density of the scalar
field is given by
%
\begin{equation}
\rho_\varphi'+3{\cal H}\rho_\varphi=-a\Gamma\rho_{\varphi},
\label{reheat1}
\end{equation}
%
where $\rho_\varphi=\varphi'^2/(2a^2)+U$, $\Gamma$ is the decay rate
of $\varphi$ and the prime denotes the derivative with respect to
$\tau$.  This has an analytic solution $\rho_\varphi\propto
a^{-3}\exp(-\Gamma t)$, where $t$ denotes cosmic time $dt=a d\tau$.
Assuming that all the decay products of the inflaton are rapidly
thermalized, the energy conservation equation of radiation density
reads
%
\begin{equation}
\rho_r'+4{\cal H}\rho_r=a\Gamma\rho_{\varphi}.\label{reheat2}
\end{equation}
%
The expansion rate of the Universe, ${\cal H}\equiv a'/a$, is
determined by the sum of the energy densities,
%
\begin{equation}
{\cal H}^2=\frac{a^2}{3M_{\rm pl}^2}(\rho_{\varphi}+\rho_r),\label{reheat3}
\end{equation}
where $M_{\rm pl}=(8\pi G)^{-1/2}$ is the reduced Planck scale.
We numerically solve these three equations, Eqs. (\ref{reheat1}) --
(\ref{reheat3}), to calculate the Hubble expansion rate in our
simulation, although they admit the following analytic solution 
\begin{align}
 \rho_\phi(t)&=\rho_\phi(t_i)\left[\frac{a(t)}{a(t_i)}\right]^{-3}
 e^{-\Gamma(t-t_i)} \\
\rho_r(t)&=\Gamma\int_{t_i}^t \left[\frac{a(t)}{a(t')}\right]^{-3}
\rho_\phi(t')dt', \label{rad}
\end{align}
which is valid if $\rho_r(t_i)$ is negligible.
For $t_i \ll t \ll \Gamma^{-1} $ (\ref{rad}) can be expressed as
\begin{equation}
 \rho_r(t)=\frac{3}{5}\Gamma t \left[\frac{a(t)}{a(t_i)}\right]^{-3}
\rho_\phi(t_i) \cong \frac{6}{5}\Gamma H M_{\rm pl}^2,
\end{equation}
with $H={\cal H}/a$.  The temperature of the Universe is therefore
given by
%
\begin{equation}
T=\left(\frac{30}{\pi^2g_*}\rho_r\right)^{\frac{1}{4}}
\simeq \left(\frac{36}{\pi^2 g_*}\Gamma H M_{\rm pl}^2\right)^{\frac{1}{4}},
\label{eq:temp}
\end{equation}
%
where $g_*$ is the effective number of relativistic degrees of
freedom.  The universe becomes radiation dominant at $t\simeq
\Gamma^{-1}$.  The temperature at this time, the reheating temperature,
is determined using this relation as
%
\begin{equation}
T_{\rm RH}\simeq
\left(\frac{10}{\pi^2g_*}\right)^{\frac{1}{4}}(\Mpl\Gamma)^{\frac{1}{2}}.
\end{equation}
%

\subsection{Global O(N) symmetric scalar field model}
We consider an N-component real scalar field
$\Phi=(\phi_1,\phi_2,...,\phi_a,...,\phi_N)$ with a Lagrangian
%
\begin{equation}
  \Lag(\Phi) = -\frac{1}{2}(\partial_\mu\Phi)^T(\partial^\mu\Phi) - V_{\rm eff}(\Phi,T),
\end{equation}
%
with a temperature-dependent effective potential \cite{Yamaguchi:1999yp}
%
\begin{equation}
V_{\rm eff}(\Phi,T) = 
 \frac{\lambda}{2}(\Phi^2 - v^2)^2 + \frac{\lambda}{3}T^2\Phi^2,
\label{eq:potential}
\end{equation}
%
where $\Phi^2 = \sum_a\phi_a^2$, $\lambda$ is the dimensionless
self-coupling of $\Phi$, and $v$ is the magnitude of the vacuum
expectation value in the true vacuum.  Throughout the paper, we take
$\lambda = 1$.  The symmetry is broken below the critical temperature
$T_c = \sqrt{3}v$.  After symmetry breaking, the scalar field acquires
a vacuum expectation value and satisfies $\Phi^2=v^2$.

The equations of motion for each component of the scalar field is
given by
%
\begin{equation}
\phi_a''(\tau,\xx) + 2\HH\phi_a'(\tau,\xx) - \nabla^2 \phi_a(\tau,\xx) = -a^2\frac{\partial V_{\rm eff}}{\partial\phi_a}.
\label{eq:phi}
\end{equation}
%
In order to set the initial condition without ambiguities we assume
that the symmetry is restored by high-temperature effects after
inflation characterized by (\ref{eq:temp}) and start simulations from
a symmetric state.  To realize the initial condition, we generate
zero-mean Gaussian random values for $\widetilde{\phi}_a(t_0,\kk)$ and
$\dot{\widetilde{\phi}}_a(t_0,\kk)$ on a discrete grid in the Fourier
space with the variance \cite{Yamaguchi:1999yp,Hiramatsu:2010yz},
%
\begin{equation}
\vol{|\widetilde{\phi_a}(t_0,\kk)|^2} = VP(\tau_0,|\kk|),\quad
\vol{|\dot{\widetilde{\phi}}_a(t_0,\kk)|^2} = VQ(\tau_0,|\kk|),
\label{eq:var}
\end{equation}
%
with
%
\begin{equation}
  P(\tau,k) = \frac{1}{\omega_k}\frac{1}{e^{\beta_T\omega_k}-1},\quad
  Q(\tau,k) = \frac{\omega_k}{e^{\beta_T\omega_k}-1},
\label{eq:PQ}
\end{equation}
%
where $\omega_k = \sqrt{k^2+m^2}$ with $m^2 = d^2V_{\rm
  eff}/d\Phi^2|_{\Phi=0}$ being the effective mass of the scalar
field, $\beta_T=1/T$ and $V=L^3$ is the comoving volume of the
simulation box.  Note that here we use the proper time $t$ instead of
$\tau$ and the dot denotes the derivative with respect to $t$.  Then,
transforming $\widetilde{\phi}_a$ and $\dot{\widetilde{\phi}}_a$ to
the real space, $\phi_a(\tau_0,\xx)$ and $\dot{\phi}_a(\tau_0,\xx)$
have the desired thermal distribution.

\subsection{Gravitational waves}
Gravitational waves are represented by a transverse-traceless
gauge-invariant metric perturbation, $h_{ij}$, in a Friedmann
Robertson-Walker background.
%
\begin{equation}
 ds^2 = a^2(\tau)\left[ -d\tau^2 + (\delta_{ij}+h_{ij})dx^idx^j \right],
\end{equation}
%
where $h_{ij}$ satisfies $h^{ij}{}_{,j}=h^{i}{}_{i}=0$.  Expanding the
Einstein equations to first order in $h_{ij}$, we obtain the equation
of motion
%
\begin{equation}
 h_{ij}''(\tau,\xx) + 2\HH h_{ij}'(\tau,\xx) - \nabla^2 h_{ij}(\tau,\xx) = \frac{2}{\Mpl^2}
   \Pi_{ij}^{\rm TT}(\tau,\xx),
\label{eq:hij}
\end{equation}
%
where the source term, $\Pi_{ij}^{\rm TT}(\tau,\xx)$ is the transverse-traceless
projection of the anisotropic stress tensor,
%
\begin{equation}
\Pi_{ij}(\tau,\xx)=\sum_a\left[\partial_i\phi_a(\tau,\xx)\partial_j\phi_a(\tau,\xx)-\frac{1}{3}\delta_{ij}
\partial_k\phi_a(\tau,\xx)\partial^k\phi_a(\tau,\xx)\right].
\end{equation}
%
The transverse-traceless
part is obtained by applying the projection operator in the momentum
space
%
\begin{equation}
  \Pi_{ij}^{\rm TT}(\tau,\kk) = \Lambda_{ij,k\ell}(\hat{k}) \Pi_{k\ell}(\tau,\kk) = \Lambda_{ij,k\ell}(\hat{k})\sum_a\{\partial_k\phi_a\partial_\ell\phi_a\}(\tau,\kk),
\end{equation}
%
with
%
\begin{eqnarray}
&& \Lambda_{ij,k\ell}(\hat{k}) = \mathcal{P}_{ik}(\hat{k})\mathcal{P}_{j\ell}(\hat{k})
  - \frac{1}{2}\mathcal{P}_{ij}(\hat{k})\mathcal{P}_{k\ell}(\hat{k}),\\
&& \mathcal{P}_{ij}(\hat{k}) = \delta_{ij} - \hat{k}_i\hat{k}_j,
\end{eqnarray}
%
where $\{\partial_k\phi_a\partial_\ell\phi_a\}(\tau,\kk)$ denotes the
Fourier transform of
$\partial_k\phi_a(\tau,\xx)\partial_\ell\phi_a(\tau,\xx)$ and
$\hat{k}=\kk/k$ \cite{Dufaux:2007pt}.

\section{Lattice simulation}
\label{sec:result}
\subsection{Set up}
In order to calculate the gravitational wave spectrum, we perform
lattice simulations by numerically evolving the scalar field and
gravitational wave on a discrete lattice. 
Introducing new variables $\psi_a$ and $\chi_{ij}$ defined as
$\phi_a=\psi_a/a$ and $h_{ij}=\Lambda_{ij,k\ell}\chi_{k\ell}/a$, we
solve the following equations
%
\begin{align}
 &\psi_a''(\tau,\xx) -\frac{a''}{a}\psi_a(\tau,\xx) - \nabla^2\psi_a(\tau,\xx) = - 2\lambda a^2\left(\Phi^2 - v^2 + \frac{T^2}{3}\right) \psi_a(\tau,\xx),\\
 &\chi_{ij}''(\tau,\xx) -\frac{a''}{a}\chi_{ij}(\tau,\xx) - \nabla^2 \chi_{ij}(\tau,\xx) = \frac{2}{\Mpl^2 a} 
\sum_a\left[\partial_i\psi_a(\tau,\xx)\partial_j\psi_a(\tau,\xx)\right].
\end{align}
%
The energy density of the gravitational waves 
is given by
%
\begin{equation}
\rho_{\rm GW}(\tau) = \frac{\Mpl^2}{4a^2} \langle h_{ij}'(\tau,\xx)h_{ij}'(\tau,\xx)\rangle_{V}
  = \frac{\Mpl^2}{4a^4} 
\frac{1}{V}\int \frac{d^3k}{(2\pi)^3}\,
[\Lambda_{ij,kl}(\hat{k})\chi_{kl}'(\tau,\kk)]^2,
\label{eq:defene_std}
\end{equation}
%
where $\langle\cdots\rangle_{V}$ represents the average over the
spatial volume and $\chi_{ij}(\tau,\kk)$ is the Fourier transform of
$\chi_{ij}(\tau,\xx)$.  It is commonly parameterized by the
dimensionless density parameter per logarithmic frequency interval,
%
\begin{equation}
  \Omega_{\rm GW}(k) \equiv \frac{1}{\rho_c}\frac{d\rho_{G\rm W}}{d\log k},
\label{eq:omega_gw}
\end{equation}
%
where $\rho_c$ is the critical density of the Universe, $\rho_c =
3\Mpl^2\HH^2/a^2$.  By substituting Eq. (\ref{eq:defene_std}) into
(\ref{eq:omega_gw}), we obtain
%
\begin{equation}
  \Omega_{\rm GW}(k,\tau)
  = \frac{k^3}{96\pi^3\HH^2a^2V}
    \int d\Omega\,\chi_{ij}'(\tau,\kk)\chi_{ij}'^*(\tau,\kk),
\end{equation}
%
where $d\Omega$ is the integral over the solid angle.  Using
$\rho_{\rm rad}\propto g_*^{-1/3} a^{-4}$ and $\rho_{\rm GW}\propto
a^{-4}$, the energy spectrum at the present time is related to that at
the end of simulation as
%
\begin{equation}
 \Omega_{{\rm GW},0}(k) = \Omega_{{\rm rad},0}\left(\frac{g_{*,0}}{g_{*,f}}\right)^{1/3}\Omega_{{\rm GW},f}(k),
\label{Eq:OmegaGW0}
\end{equation}
in the plateau region where gravitational waves are generated after the
entropy production from the inflaton has been terminated.
%
Here the subscript $0$ denotes the value at the present time and $f$
denotes the value at the end of simulation.  Using entropy
conservation, $g_*T^3\propto a^{-3}$, the physical wavenumber at the
end of simulation $k_{{\rm phys},f}$ is related to the frequency today
as
%
\begin{equation}
 f_0 = \frac{k_{{\rm phys},f}}{2\pi} \frac{a_f}{a_0} = \frac{k_{{\rm phys},f}}{2\pi}\left(\frac{g_{*,0}}{g_{*,f}}\right)^{1/3}\left(\frac{T_0}{T_f}\right).
\label{Eq:f0}
\end{equation}
%

\begin{figure}
 \begin{center}
   \includegraphics[width=0.48\textwidth]{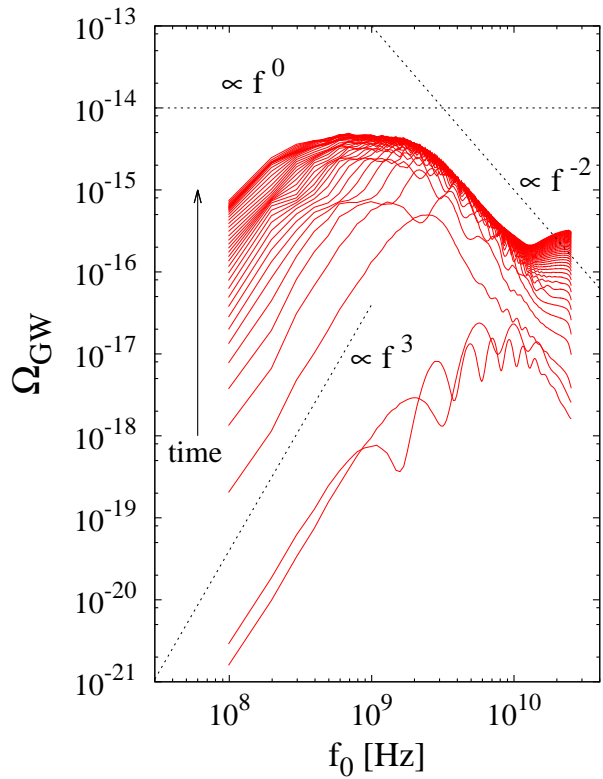} \caption{\label{Fig:1}
     The gravitational wave spectrum shown in terms of the present
     gravitational frequency $f_0$[Hz].  Time evolution is shown from
     bottom to top.  The number of the field component is $N=4$.  The
     decay rate is taken to be $\Gamma = 0.2v$, which corresponds to
     $T_{\rm RH}=2.2\times 10^{16}$GeV.  }
 \end{center}
\end{figure}

\begin{figure}
 \begin{center}
   \includegraphics[width=0.48\textwidth]{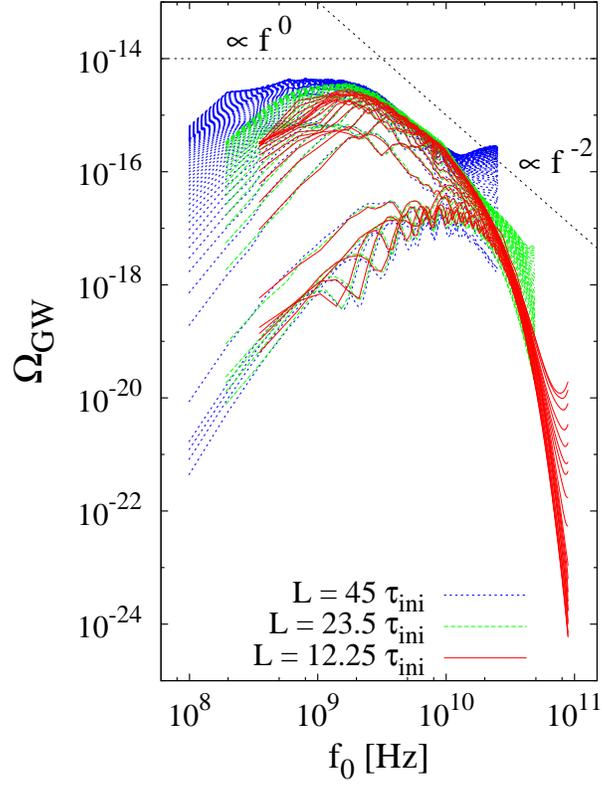} \caption{\label{Fig:2}
     The gravitational wave spectra for different simulation box
     sizes.  The number of the field component is fixed to be $N=4$. }
 \end{center}
\end{figure}

\begin{figure}
 \begin{center}
   \includegraphics[width=0.48\textwidth]{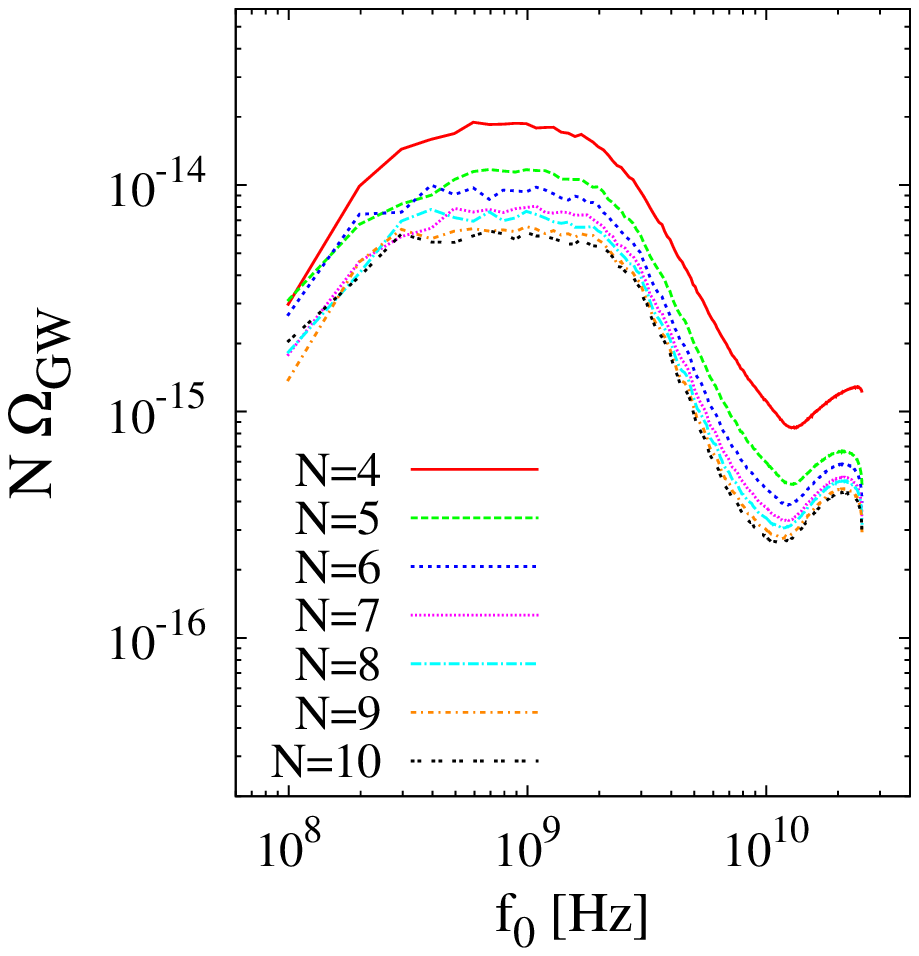} \caption{\label{Fig:3}
     The gravitational wave spectra for different values of $N$.  The
     vertical axis is multiplied by $N$ to test the analytically
     predicted dependence of $\Omega_{\rm GW}\propto 1/N$.  }
 \end{center}
\end{figure}

We perform simulations with $512^3$ lattices assuming $v=9\times
10^{-4}\mpl$, $g_{*,f}=1000$ and $\Gamma = 0.2v$, which corresponds to
$T_{\rm RH}=2.2\times 10^{16}$GeV.  The simulation starts before the
phase transition.  We take the initial scale factor to be $a_{\rm
  init}=1$ and the initial Hubble size to be ${\cal H}_{\rm
  init}=1.1v$, which relates to the conformal time as $\tau_{\rm
  init}= 2/{\cal H}_{\rm init} = 1.8/v$.  The simulation stops when
the comoving Hubble radius becomes a half of the box size, ${\cal
  H}_{\rm end}^{-1}=L/2$.  Then the result is converted to the present
value using Eqs. (\ref{Eq:OmegaGW0}) and (\ref{Eq:f0}) with
$g_{*,0}=3.36$, $\Omega_{{\rm rad},0}h^2=4.15\times 10^{-5}$ and
$T_0=2.725$K, where $h=0.7$ is the reduced Hubble parameter.

Note that we take a value of $v$ larger than the current CMB
constraint and a nonstandard value of $g_*$ in our simulations.
Furthermore the initial value of the Hubble parameter at the beginning
of simulations also exceeds the constraints on the Hubble parameter
imposed by the tensor perturbations generated during standard
inflation, although we can evade such a constraint if we adopt an
inflation model in which the null energy condition is violated
\cite{Kobayashi:2015gga}.  The reason why we choose such nonstandard
values of $v$ an $g_*$ is that, for small values of $v$ and $g_{*}$,
gravitational waves generated by self-ordering scalar fields after the
phase transition are contaminated by those generated from the thermal
fluctuations of the scalar fields before the transition.  In order to
suppress this contamination and focus on the effect from reheating, we
use relatively large values of $v$ and $g_*$.  In fact, large $v$
enhances the power of gravitational wave spectrum from self-ordering
with the dependence of $\Omega_{\rm GW}\propto v^4$, and 
large $g_*$ reduces the initial temperature and gives smaller thermal
fluctuations before the phase transition (see Eqs. (\ref{eq:temp}) and
(\ref{eq:PQ}) for the temperature dependence of the thermal
fluctuations).

Once simulations with such parameters are performed we can obtain
results with other parameter values using the above mentioned scaling
law $\Omega_{\rm GW} \propto v^4$ as well as other scaling discussed below.

\subsection{Results}

Figure 1 shows time evolution of the gravitational wave spectrum.  We
see that higher-frequency modes come inside the horizon earlier and
the gravitational wave is generated soon after the mode comes into the
horizon.  The modes outside the horizon have frequency dependence of
$f^3$, which matches the analytical prediction of
Refs. \cite{JonesSmith:2007ne,Fenu:2013tea}.  Note that the $f^3$
dependence remains in the final spectrum at the lowest frequencies,
since we stop the simulation when the Hubble radius becomes half the
box size.  However, if we could trace the time evolution longer, the
$f^0$ dependence would continue toward the lower frequencies.  For the
modes inside the horizon at the end of simulation, we see the
frequency dependence of the spectrum changes from $f^{-2}$ to $f^0$
because of the transition from the matter-dominated to the
radiation-dominated phase.

The bump seen at the highest frequencies in the final regime of the
simulation is an artifact due to the finite resolution (see also
Fig. \ref{Fig:2}).  Gravitational waves are not produced if the mode
is already inside the horizon at the time of the phase transition.  In
Fig. \ref{Fig:2}, we show spectra for smaller box sizes which 
give a
better resolution.  We confirm that the spectrum represented in red
keeps the initial shape and thus damps exponentially at high
frequencies where the bumps appear in the computations with a worse
resolution (blue and green).

Figure \ref{Fig:3} shows the spectra in the case with different
numbers of the scalar field components.  Note that the vertical axis
is the power multiplied by $N$.  The amplitude of the spectrum has the
same dependence with the analytical prediction $\Omega_{\rm GW}\propto
1/N$ \cite{JonesSmith:2007ne,Fenu:2013tea} for large $N$.  Contrarily,
we find the extra power for small $N$, which would imply the breakdown
of the large $N$ approximation used in the analytical predictions.
This point has been studied in Ref.  \cite{Figueroa:2012kw} and our
result is consistent with them.

\subsection{Comparison with inflationary gravitational wave spectrum}
\begin{figure}
 \begin{center}
   \includegraphics[width=0.48\textwidth]{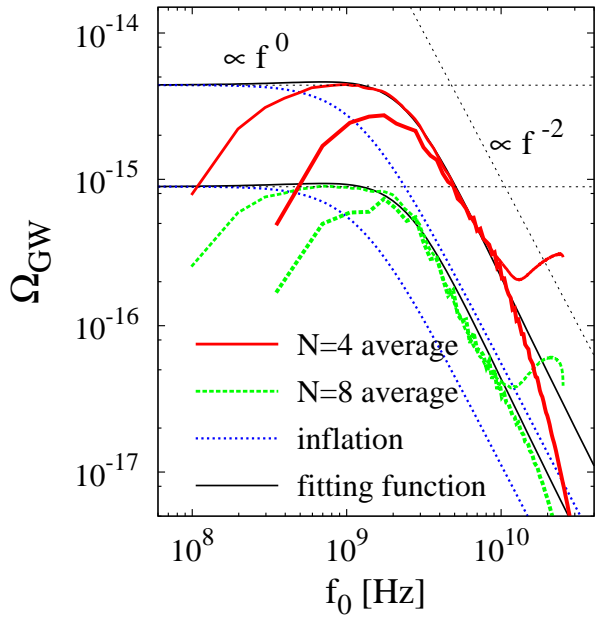} 
   \caption{\label{Fig:4}
     The gravitational wave background spectra compared with that from
     inflation.  
      The red solid lines represent the case of $N=4$ and the
       green dashed lines represent $N=8$, while blue dotted lines are
       the spectra from inflation calculated assuming the same
       reheating temperature.  The decay rate is taken to be $\Gamma =
       0.2v$ for all cases, which corresponds to $T_{\rm RH}=2.2\times
       10^{16}$GeV.  For the spectra of $O(N)$ phase transition, we
       show the averaged values of 20 realizations of simulations with
       a large box size $L=45\tau_{\rm ini}$ and a small box size
       $L=12.25\tau_{\rm ini}$. For the inflationary gravitational
       waves, we assume no tilt of the spectrum and the amplitude is
       tuned to be the same as that of $O(N)$ phase transition spectra
       at the frequencies where the spectrum has the $f^0$
       dependence.  We also show the fitting function, given by
     Eqs. (\ref{Eq:OGWparametrize}) and (\ref{Eq:T2self}).  }
 \end{center}
\end{figure}

In Fig. \ref{Fig:4}, we compare the spectrum with that from inflation.
For the spectrum from the $O(N)$ phase transition, we show the
spectrum obtained by averaging over 20 realizations of simulation for
each combination of parameters.  As explained before, the damping of
the power seen at the low frequencies is because of the limitation in
the simulation time.  The $f^0$ dependence therefore should continue
to the lower frequencies.  For the spectrum from inflation, we
normalized the amplitude at low frequencies and take the value to be
the same as the case of the $O(N)$ phase transition for comparison.
We also assume the same reheating temperature for both cases.

Comparing the two spectra, we find that there is a difference in the
position of the spectral bend.  This is because gravitational waves
are continuously generated even after the inward horizon crossing
until each $k$ mode of scalar field fluctuations is homogenized.  We
also find that the sharpness of the transition from $f^{-2}$ to $f^0$
is slightly different around the bend.

Let us introduce a fitting formula to describe the shape of the
spectrum caused by reheating and approximate the spectrum as
\begin{equation}
\Omega_{\rm GW}(f) = \Omega_{\rm GW,A}T^2(x_R), 
  \label{Eq:OGWparametrize}
\end{equation}
where $\Omega_{\rm GW,A}$ is the normalization of the spectral
amplitude and $T^2(x_R)$ is the transfer function which describes the
transition from $f^0$ to $f^{-2}$.  For inflation, the transfer
function has been found as \cite{Kuroyanagi:2014nba}
\begin{equation}
T^2_{\rm inflation}(x_R)  = \left(1 - 0.22 x_R^{1.5} + 0.65 x_R^2\right)^{-1},
\label{Eq:T2inf}
\end{equation}
where $x_R=f/f_R$ and $f_R = 0.26 \left( g_{\ast s} (T_R)/106.75
\right)^{1/6} \left( T_R/10^7~{\rm GeV} \right)$Hz.  For the $O(N)$
phase transition, we find that the spectrum is well described by
the transfer function
\begin{equation}
T^2_{O(N)}(x_{R'})  = \left(1 - 0.6 x_{R'}^{1.5} + 0.65 x_{R'}^2\right)^{-1},
\label{Eq:T2self}
\end{equation}
with $x_{R'}=f/(1.7f_R)$.  Note the factor $1.7$ in front of $f_R$ in
the definition of $x_{R'}$ represents the difference in the position
of the spectral bend.  If we tried to determine the reheating
temperature by the spectral bend without knowing the origin of
gravitational waves, we might make a wrong measurement deviated from
the true value by 70\%.  Note also that the coefficient of the second
term in the transfer function is changed from $0.22$ to $0.6$, which
is a parameter to determine the sharpness of the spectral
transition\footnote{Strictly speaking, the transfer function has a
  weak dependence on $N$.  For example, this parameter takes 0.6 for
  $N=4$ but is closer to 0.7 for $N=8$. }.  In Fig. \ref{Fig:4}, we
also show the comparison between the simulation results and the
spectra generated using the fitting formula.

\subsection{Detectability in future experiments}
\begin{figure}
 \begin{center}
   \includegraphics[width=0.9\textwidth]{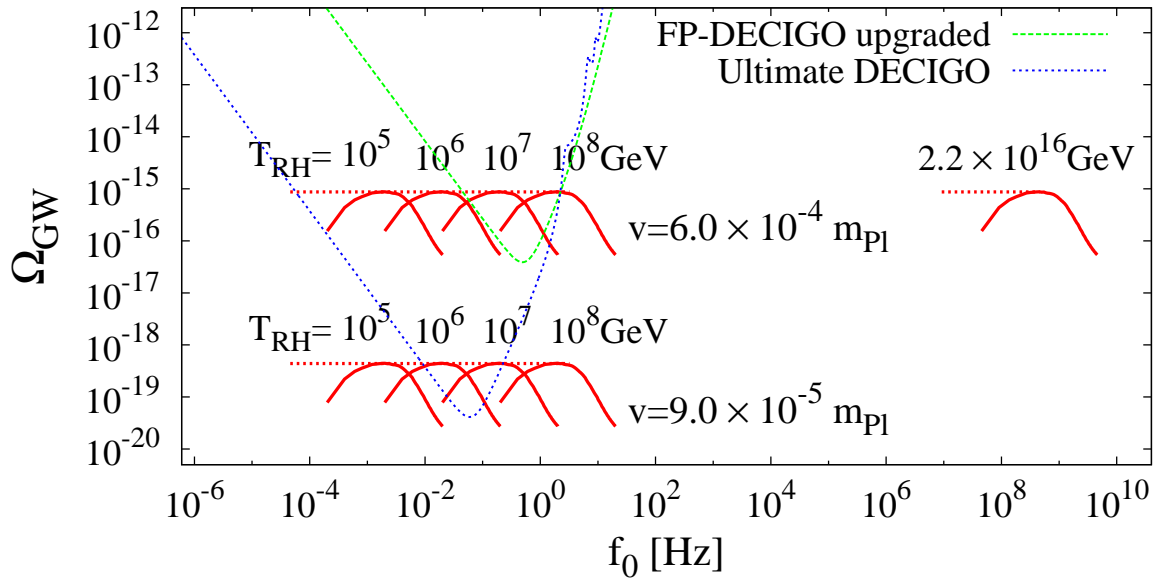} \caption{\label{Fig:5}
     Comparison with the sensitivity curves of future experiments.
     Gravitational wave spectra are plotted for different reheating
     temperatures and different values of $v$ with $g_\ast =106.75$.
     Here we use the result of $N=4$. }
 \end{center}
\end{figure}
Let us discuss the detectability of the reheating signature by future
direct detection experiments.  So far, we have shown the case with an
unrealistically high reheating temperature $T_{\rm RH}=2.2\times
10^{16}$GeV due to the limitation of the simulation time and
resolution.  For lower reheating temperatures, although we cannot
follow the whole evolution of the scalar field from the phase
transition to the completion of reheating, the gravitational wave
spectrum can be rescaled by just changing the frequency by $f_0\propto
g_*^{1/6}T_{\rm RH}$ for different reheating temperatures
\cite{Seto:2003kc,Nakayama:2008wy}.  Also, for different values of
vacuum expectation value of the scalar field, the amplitude scales as
$\Omega_{\rm GW}\propto v^4$ \cite{JonesSmith:2007ne,Fenu:2013tea}.

Using the dependence on $T_{\rm RH}$ and $v$, we show the spectra for
different reheating temperatures and different vacuum expectation
values in Fig. \ref{Fig:5}, comparing with the sensitivity curves of
the future satellite-type experiments such as DECIGO
\cite{Kawamura:2011zz}, BBO \cite{bbo} and Ultimate DECIGO
\cite{Seto:2001qf} (calculated assuming 10-year observation time).
The noise curve titled as FP-DECIGO (Fabry-P\'erot-type DECIGO) is the
upgraded version from the original FP-DECIGO, whose sensitivity is
improved about three times to remove all the foreground contamination
from neutron star binaries \cite{Yagi:2011wg}.  Ultimate-DECIGO is the
experiment which has the ideal sensitivity limited only by quantum
noises.

\begin{figure}
 \begin{center}
   \includegraphics[width=0.48\textwidth]{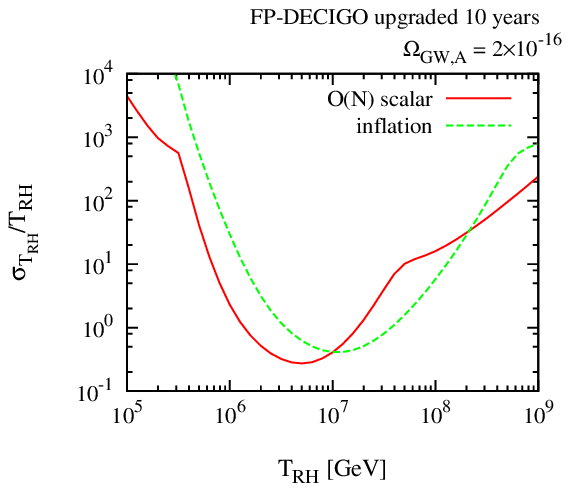}
   \includegraphics[width=0.48\textwidth]{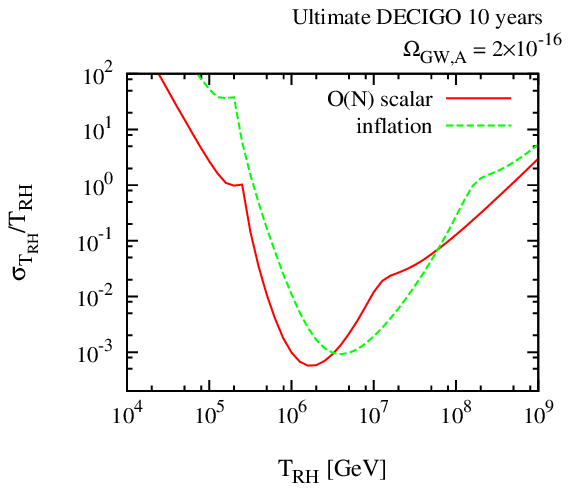}\caption{\label{Fig:6}
     The marginalized 1$\sigma$ uncertainty in $T_{\rm RH}$ as a
     function of $T_{\rm RH}$ for upgraded FP-DECIGO (left panel) and
     Ultimate DECIGO (right panel).  The red solid line represents the
     case of $O(N)$ phase transition and the green dashed line shows
     the case of inflation.  The case with $\Omega_{\rm GW, A}=2\times
     10^{-16}$ at the plateau region is shown for illustration, but
     the vertical axis simply scales as $\sigma_{T_{\rm RH}}\propto
     \Omega_{\rm GW, A}^{-1}$.  }
 \end{center}
\end{figure}

In Fig. \ref{Fig:6}, we show how accurately the reheating temperature
can be determined when the spectral shape of reheating is measured by
the future experiments.  We calculate Fisher matrices using the
parameterization of Eq. (\ref{Eq:OGWparametrize}) and estimate the
expected errors on both $\Omega_{\rm GW, A}$ and $T_{\rm RH}$ assuming
the noise designs of FP-DECIGO and Ultimate-DECIGO.  In the
calculation, we do not use the information in $f<0.1$Hz, which may be
contaminated by foreground noise from white dwarf binaries.  For the
details of the calculation method of the Fisher matrix and the noise
curves, see Ref. \cite{Kuroyanagi:2014qza}.

The curves shown in the figures are the 1$\sigma$ errors on $T_{\rm
  RH}$ marginalized over the other parameter $\Omega_{\rm GW, A}$.
Within the range of $T_{\rm RH}$ where $\sigma_{T_{\rm RH}}/T_{\rm
  RH}< 1$, one may expect that the reheating temperature can be
determined with a certain level of accuracy.  We also show the case of
inflation for comparison.  For inflationary gravitational waves, we
find that FP-DECIGO is the most sensitive at $T_{\rm RH}\sim 10^7$GeV,
\footnote{Note that the sensitive frequency range for inflation is
  slightly different from the results in Ref.
  \cite{Kuroyanagi:2014qza}.  This is because we assume different
  value of $g_*$ and also because we neglect the tilt of the spectrum
  in this paper.}  while it has a better sensitivity at slightly lower
reheating temperature in the case of the $O(N)$ phase transition.
This is because the position of the spectral bend is different
depending on the origin, as has been seen in Fig. \ref{Fig:4}.  This
difference would cause an overestimate/underestimate of the reheating
temperature by a factor of $1.7$, if one estimates the reheating
temperature with assuming an incorrect origin of the gravitational
wave background.

\begin{figure}
 \begin{center}
   \includegraphics[width=0.48\textwidth]{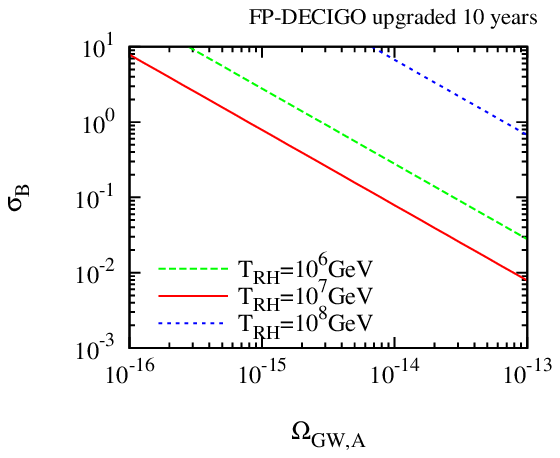}
   \includegraphics[width=0.48\textwidth]{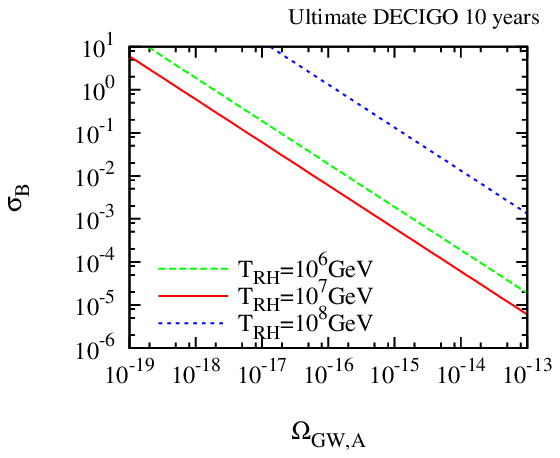}\caption{\label{Fig:7}
     The marginalized 1$\sigma$ uncertainty in the coefficient of the
     second term in the transfer function $B$, which would enable us
     to discriminate the origin of the gravitational wave background
     if $\sigma_B<0.1$.  We show the result as a function of the
     normalization amplitude $\Omega_{\rm GW, A}$ for upgraded
     FP-DECIGO (left panel) and Ultimate DECIGO (right panel).  Three
     lines represent different fiducial values of $T_{\rm RH}$.  }
 \end{center}
\end{figure}

Finally, we discuss whether it is possible to distinguish the
gravitational wave background of the $O(N)$ phase transition from that
of inflation.  As seen in Eqs. (\ref{Eq:T2inf}) and (\ref{Eq:T2self}),
both of the transfer functions have the form of $T^2(x_R) = (1 - B
x_R^{1.5} + 0.65 x_R^2)^{-1}$.  The difference of origin arises in the
coefficient of the second term $B$, which is $0.22$ for inflation and
about $0.6$ for $O(N)$ phase transition.  Therefore this parameter may help
us to identify the origin of the observed gravitational waves if the
value is precisely measured.  Here, we perform the Fisher analysis by
adding $B$ as an additional free parameter.  In Fig. \ref{Fig:7}, we
show the expected 1$\sigma$ error on $B$ marginalized over
$\Omega_{\rm GW, A}$ and $T_{\rm RH}$.  The fiducial value of $B$ is
taken as $0.6$.  We see the error on $B$ becomes smaller for larger
normalization amplitude $\Omega_{\rm GW, A}$, because it corresponds
to signal detection with a high signal-to-noise ratio.  Since we need
to measure the difference between $B=0.22$ and $0.6$, we may expect to
specify the origin if $B$ is determined with the accuracy of
$\sigma_B<0.1$.  To achieve this accuracy, in the case where the
reheating temperature is $T_{\rm RH}=10^7$GeV, the amplitude of the
gravitational wave should be larger than $\Omega_{\rm GW, A}=8\times
10^{-15}$ for FP-DECIGO, and $\Omega_{\rm GW, A}=6\times 10^{-18}$ for
Ultimate DECIGO.


\section{Conclusion}
\label{sec:conclusion}
We have investigated how the effect of reheating appears in the
gravitational wave background spectrum of the global $O(N)$ phase
transition.  Using numerical lattice simulations, we showed the
power-law dependence of the spectrum changes from $f^{-2}$ to $f^0$,
which is an imprint of the change of the Hubble expansion rate from
inflaton-oscillation dominated to radiation dominated regimes at the
end of reheating.  We also compared our result with the spectrum of
inflation-produced gravitational waves and found differences in the
position of the reheating signature and in the shape of the spectrum.
By introducing a fitting function of the spectral shape, we performed
a Fisher analysis to discuss whether one can determine the reheating
temperature by detecting the gravitational waves from the $O(N)$ phase
transition with future experiments.  We have also investigated whether
one can distinguish the origin between $O(N)$ phase transition and
inflation by detecting the small difference in the spectral shape alone.

Observation of gravitational wave provides a unique opportunity to
probe the very early epoch of the Universe.  Inflation is one of the
strong candidates as a generation mechanism of gravitational waves in
the early Universe, which could be used as a tool to extract
information on the thermal history after inflation.  At the same time,
we should always keep in mind that there may be alternative ways to
probe it using gravitational waves from different origins.  The change
of the Hubble expansion rate in general affects the evolution of
gravitational waves through the Hubble expansion term in the evolution
equation.  Gravitational waves from a global $O(N)$ phase transition
is a good example which clearly contains the effect in its
scale-invariant spectrum.  Although this paper has focused on the
effect under the background of the conventional reheating model, it
could be used to test any type of mechanism in the early Universe
which induces a different behavior of the Hubble expansion rate.

\begin{acknowledgements}
  This work is partly supported by Career Development Project for
  Researchers of Allied Universities (SK).  This work was partially
  supported by JSPS KAKENHI Grant Numbers 23340058(JY) and
  15H02082. This work was supported in part by MEXT SPIRE and JICFuS
  (TH).
\end{acknowledgements}

\bibliographystyle{apsrev}

\begin{thebibliography}{}

\bibitem{Harry:2010zz} 
  G.~M.~Harry [LIGO Scientific Collaboration],
  Class.\ Quant.\ Grav.\  {\bf 27}, 084006 (2010).

\bibitem{Accadia:2011zzc} 
  T.~Accadia, F.~Acernese, F.~Antonucci, P.~Astone, G.~Ballardin, F.~Barone, M.~Barsuglia and A.~Basti {\it et al.},
  Class.\ Quant.\ Grav.\  {\bf 28}, 114002 (2011).

\bibitem{Somiya:2011np} 
  K.~Somiya [KAGRA Collaboration],
  Class.\ Quant.\ Grav.\  {\bf 29}, 124007 (2012)
  [arXiv:1111.7185 [gr-qc]].

\bibitem{AmaroSeoane:2012km} 
  P.~Amaro-Seoane, S.~Aoudia, S.~Babak, P.~Binetruy, E.~Berti, A.~Bohe, C.~Caprini and M.~Colpi {\it et al.},
  arXiv:1201.3621 [astro-ph.CO].

\bibitem{AmaroSeoane:2012je} 
  P.~Amaro-Seoane, S.~Aoudia, S.~Babak, P.~Binetruy, E.~Berti, A.~Bohe, C.~Caprini and M.~Colpi {\it et al.},
  Class.\ Quant.\ Grav.\  {\bf 29}, 124016 (2012)
  [arXiv:1202.0839 [gr-qc]].

\bibitem{Seto:2001qf} 
  N.~Seto, S.~Kawamura and T.~Nakamura,
  Phys.\ Rev.\ Lett.\  {\bf 87}, 221103 (2001)
  [astro-ph/0108011].


\bibitem{Kawamura:2011zz} 
  S.~Kawamura, M.~Ando, N.~Seto, S.~Sato, T.~Nakamura, K.~Tsubono, N.~Kanda and T.~Tanaka {\it et al.},
  Class.\ Quant.\ Grav.\  {\bf 28}, 094011 (2011).

\bibitem{bbo}
S. Phinney {\it et al.}, 
{\it The big bang observer: direct detection of gravitational waves from the birth of the Universe to the present}, 
NASA Mission Concept Study. 

\bibitem{Krauss:1991qu} 
  L.~M.~Krauss,
  Phys.\ Lett.\ B {\bf 284}, 229 (1992).

\bibitem{Kamionkowski:1996zd} 
  M.~Kamionkowski, A.~Kosowsky and A.~Stebbins,
  Phys.\ Rev.\ Lett.\  {\bf 78}, 2058 (1997)
  [astro-ph/9609132].

\bibitem{Zaldarriaga:1996xe} 
  M.~Zaldarriaga and U.~Seljak,
  Phys.\ Rev.\ D {\bf 55}, 1830 (1997)
  [astro-ph/9609170].

\bibitem{Planck:2006aa} 
  J.~Tauber {\it et al.}  [Planck Collaboration],
  astro-ph/0604069.

\bibitem{Hanson:2013hsb} 
  D.~Hanson {\it et al.}  [SPTpol Collaboration],
  Phys.\ Rev.\ Lett.\  {\bf 111}, 141301 (2013)
  [arXiv:1307.5830 [astro-ph.CO]].

\bibitem{Ade:2014afa} 
  P.~A.~R.~Ade {\it et al.}  [ The POLARBEAR Collaboration],
  arXiv:1403.2369 [astro-ph.CO].

\bibitem{Hobbs:2009yy} 
  G.~Hobbs, A.~Archibald, Z.~Arzoumanian, D.~Backer, M.~Bailes, N.~D.~R.~Bhat, M.~Burgay and S.~Burke-Spolaor {\it et al.},
  Class.\ Quant.\ Grav.\  {\bf 27}, 084013 (2010)
  [arXiv:0911.5206 [astro-ph.SR]].

\bibitem{vanHaasteren:2011ni} 
  R.~van Haasteren, Y.~Levin, G.~H.~Janssen, K.~Lazaridis, M.~Kramer, B.~W.~Stappers, G.~Desvignes and M.~B.~Purver {\it et al.},
  Mon.\ Not.\ Roy.\ Astron.\ Soc.\  {\bf 414}, no. 4, 3117 (2011)
  [Erratum-ibid.\  {\bf 425}, no. 2, 1597 (2012)]
  [arXiv:1103.0576 [astro-ph.CO]].

\bibitem{Demorest:2012bv} 
  P.~B.~Demorest, R.~D.~Ferdman, M.~E.~Gonzalez, D.~Nice, S.~Ransom, I.~H.~Stairs, Z.~Arzoumanian and A.~Brazier {\it et al.},
  Astrophys.\ J.\  {\bf 762}, 94 (2013)
  [arXiv:1201.6641 [astro-ph.CO]].

\bibitem{Manchester:2012za} 
  R.~N.~Manchester, G.~Hobbs, M.~Bailes, W.~A.~Coles, W.~van Straten, M.~J.~Keith, R.~M.~Shannon and N.~D.~R.~Bhat {\it et al.},
  arXiv:1210.6130 [astro-ph.IM].

\bibitem{JonesSmith:2007ne} 
  K.~Jones-Smith, L.~M.~Krauss and H.~Mathur,
  Phys.\ Rev.\ Lett.\  {\bf 100}, 131302 (2008)
  [arXiv:0712.0778 [astro-ph]].

\bibitem{Fenu:2009qf} 
  E.~Fenu, D.~G.~Figueroa, R.~Durrer and J.~Garcia-Bellido,
  JCAP {\bf 0910}, 005 (2009)
  [arXiv:0908.0425 [astro-ph.CO]].

\bibitem{Giblin:2011yh} 
  J.~T.~Giblin, Jr., L.~R.~Price, X.~Siemens and B.~Vlcek,
  JCAP {\bf 1211}, 006 (2012)
  [arXiv:1111.4014 [astro-ph.CO]].

\bibitem{Figueroa:2012kw} 
  D.~G.~Figueroa, M.~Hindmarsh and J.~Urrestilla,
  Phys.\ Rev.\ Lett.\  {\bf 110}, no. 10, 101302 (2013)
  [arXiv:1212.5458 [astro-ph.CO]].

\bibitem{Durrer:1998rw} 
  R.~Durrer, M.~Kunz and A.~Melchiorri,
  Phys.\ Rev.\ D {\bf 59}, 123005 (1999)
  [astro-ph/9811174].

\bibitem{GarciaBellido:2010if} 
  J.~Garcia-Bellido, R.~Durrer, E.~Fenu, D.~G.~Figueroa and M.~Kunz,
  Phys.\ Lett.\ B {\bf 695}, 26 (2011)
  [arXiv:1003.0299 [astro-ph.CO]].

\bibitem{Fenu:2013tea} 
  E.~Fenu, D.~G.~Figueroa, R.~Durrer, J.~Garcia-Bellido and M.~Kunz,
  arXiv:1311.3225 [astro-ph.CO].

\bibitem{Adshead:2009bz} 
  P.~Adshead and E.~A.~Lim,
  Phys.\ Rev.\ D {\bf 82}, 024023 (2010)
  [arXiv:0912.1615 [astro-ph.CO]].

\bibitem{Figueroa:2010zx} 
  D.~G.~Figueroa, R.~R.~Caldwell and M.~Kamionkowski,
  Phys.\ Rev.\ D {\bf 81}, 123504 (2010)
  [arXiv:1003.0672 [astro-ph.CO]].

\bibitem{Amin:2014ada} 
  M.~A.~Amin and D.~Grin,
  arXiv:1405.1039 [astro-ph.CO].

\bibitem{Baumann:2009mq} 
  D.~Baumann and M.~Zaldarriaga,
  JCAP {\bf 0906}, 013 (2009)
  [arXiv:0901.0958 [astro-ph.CO]].

\bibitem{Adshead:2009bz} 
  P.~Adshead and E.~A.~Lim,
  Phys.\ Rev.\ D {\bf 82}, 024023 (2010)
  [arXiv:0912.1615 [astro-ph.CO]].

\bibitem{Krauss:2010df} 
  L.~M.~Krauss, K.~Jones-Smith, H.~Mathur and J.~Dent,
  Phys.\ Rev.\ D {\bf 82}, 044001 (2010)
  [arXiv:1003.1735 [astro-ph.CO]].

\bibitem{Ade:2013xla} 
  P.~A.~R.~Ade {\it et al.}  [Planck Collaboration],
  Astron.\ Astrophys.\  {\bf 571}, A25 (2014)
  [arXiv:1303.5085 [astro-ph.CO]].

\bibitem{Urrestilla:2007sf} 
  J.~Urrestilla, N.~Bevis, M.~Hindmarsh, M.~Kunz and A.~R.~Liddle,
  JCAP {\bf 0807}, 010 (2008)
  [arXiv:0711.1842 [astro-ph]].

\bibitem{Dent:2014rga} 
  J.~B.~Dent, L.~M.~Krauss and H.~Mathur,
  arXiv:1403.5166 [astro-ph.CO].

\bibitem{Durrer:2014raa} 
  R.~Durrer, D.~G.~Figueroa and M.~Kunz,
  arXiv:1404.3855 [astro-ph.CO].

\bibitem{Lizarraga:2014xza} 
  J.~Lizarraga, J.~Urrestilla, D.~Daverio, M.~Hindmarsh, M.~Kunz and A.~R.~Liddle,
  Phys.\ Rev.\ D {\bf 90}, no. 10, 103504 (2014)
  [arXiv:1408.4126 [astro-ph.CO]].

\bibitem{Seto:2003kc} 
  N.~Seto and J.~Yokoyama,
  J.\ Phys.\ Soc.\ Jap.\  {\bf 72}, 3082 (2003)
  [gr-qc/0305096].

\bibitem{Boyle:2005se} 
  L.~A.~Boyle and P.~J.~Steinhardt,
  Phys.\ Rev.\ D {\bf 77}, 063504 (2008)
  [astro-ph/0512014].

\bibitem{Nakayama:2008ip} 
  K.~Nakayama, S.~Saito, Y.~Suwa and J.~Yokoyama,
  Phys.\ Rev.\ D {\bf 77}, 124001 (2008)
  [arXiv:0802.2452 [hep-ph]];

\bibitem{Nakayama:2008wy} 
K.~Nakayama, S.~Saito, Y.~Suwa and J.~Yokoyama,
  JCAP {\bf 0806}, 020 (2008)
  [arXiv:0804.1827 [astro-ph]].

\bibitem{Nakayama:2009ce} 
  K.~Nakayama and J.~Yokoyama,
  JCAP {\bf 1001}, 010 (2010)
  [arXiv:0910.0715 [astro-ph.CO]].

\bibitem{Kuroyanagi:2010mm} 
  S.~Kuroyanagi, T.~Chiba and N.~Sugiyama,
  Phys.\ Rev.\ D {\bf 83}, 043514 (2011)
  [arXiv:1010.5246 [astro-ph.CO]].

\bibitem{Kuroyanagi:2011fy} 
  S.~Kuroyanagi, K.~Nakayama and S.~Saito,
  Phys.\ Rev.\ D {\bf 84}, 123513 (2011)
  [arXiv:1110.4169 [astro-ph.CO]].

\bibitem{Kuroyanagi:2013ns} 
  S.~Kuroyanagi, C.~Ringeval and T.~Takahashi,
  Phys.\ Rev.\ D {\bf 87}, 083502 (2013)
  [arXiv:1301.1778 [astro-ph.CO]].

\bibitem{Yamaguchi:1999yp}
  M.~Yamaguchi,
  Phys.\ Rev.\ D {\bf 60} (1999) 103511
  [hep-ph/9907506].

\bibitem{Hiramatsu:2010yz} 
  T.~Hiramatsu, M.~Kawasaki and K.~'i.~Saikawa,
  JCAP {\bf 1005}, 032 (2010)
  [arXiv:1002.1555 [astro-ph.CO]].

\bibitem{Dufaux:2007pt} 
  J.~F.~Dufaux, A.~Bergman, G.~N.~Felder, L.~Kofman and J.~-P.~Uzan,
  Phys.\ Rev.\ D {\bf 76}, 123517 (2007)
  [arXiv:0707.0875 [astro-ph]].
\bibitem{Kobayashi:2015gga} 
  T.~Kobayashi, M.~Yamaguchi and J.~Yokoyama,
  JCAP {\bf 1507}, no. 07, 017 (2015)
  [arXiv:1504.05710 [hep-th]].
\bibitem{Kuroyanagi:2014nba} 
  S.~Kuroyanagi, T.~Takahashi and S.~Yokoyama,
  arXiv:1407.4785 [astro-ph.CO].

\bibitem{Yagi:2011wg} 
  K.~Yagi and N.~Seto,
  Phys.\ Rev.\ D {\bf 83}, 044011 (2011)
  [arXiv:1101.3940 [astro-ph.CO]].

\bibitem{Kuroyanagi:2014qza} 
  S.~Kuroyanagi, K.~Nakayama and J.~Yokoyama,
  PTEP {\bf 2015}, no. 1, 013E02 (2015)
  [arXiv:1410.6618 [astro-ph.CO]].


\end{thebibliography}

\end{document}